\documentclass{aa}
\usepackage{graphicx}
\usepackage{txfonts}
\def\ms{\,m\,s$^{-1}$}         
\def\cms{\,cm\,s$^{-1}$}       
\def\m2s2{\,m$^{2}$\,s$^{-2}$} 

\begin{document}
   \title{Asteroseismology of the planet-hosting star $\mu$ Arae. \\
   I. The acoustic spectrum\thanks{Based 
   on observations collected with the HARPS spectrographs at the 3.6-m telescope 
	   (La Silla Observatory, ESO, Chile: program 073.D-0578)}
	 }  


   \author{F. Bouchy\inst{1,2}, M. Bazot\inst{3}, N.C. Santos\inst{4,5},
           S. Vauclair\inst{3}, \and D. Sosnowska\inst{5}
          }

   \offprints{\email{Francois.Bouchy@oamp.fr}}

   \institute{Laboratoire d'Astrophysique de Marseille, 
               Traverse du Siphon, 13013 Marseille, France
         \and
	      Observatoire de Haute Provence,
	      04870 St Michel l'Observatoire, France\\
             \email{Francois.Bouchy@oamp.fr}              
         \and
	      Universit\'e Paul Sabatier, Observatoire Midi-Pyr\'en\'ees, 14 av. E.
	      Belin, 31400 Toulouse, France  
         \and
	     Lisbon Observatory, Tapada da Ajuda, 1349-018 Lisboa, Portugal
	 \and
              Observatoire de Gen\`eve, 51 ch. des Maillettes, 1290 Sauverny, Switzerland	        
            }

   \date{Received ; accepted }

   \authorrunning{Bouchy et al.}
   \titlerunning{The acoustic spectrum of $\mu$ Arae}

   \abstract{We present HARPS spectroscopy of $\mu$~Arae (HD160691) performed for studying the 
origin of the metallicity excess in this planet-hosting stars. The asteroseismologic campaign led 
to the previously reported discovery of a 14 earth mass planetary companion (Santos et al. \cite{santos04b}). 
The present analysis reinforces this interpretation by excluding other possible processes 
for explaining the observed Doppler variation and 
leads to the identification of up to 43 p-mode oscillations with $l$=0-3, frequencies 
in the range 1.3-2.5 mHz and amplitudes in the range 10-40 {\cms}.  

   }

   \maketitle
%

\section{Introduction}

One particularly important fact that is helping astronomers to understand the
mechanisms of planetary formation has to do with the planet-hosting stars themselves. 
These stars were found to be particularly metal-rich, i.e. they have, on the
average, a metal content higher, by $\sim$0.2 dex, than the one found in field stars without 
detected planetary companions (e.g. Gonzalez \cite{gonzalez98}; Santos et al. 
\cite{santos01}, \cite{santos03}, \cite{santos04a}). 

The impact of this fact is enormous regarding the understanding of the
processes of planetary formation and evolution. If stars with planets were 
formed out of more-than-average metallic material, then we might conclude that 
the metallicity is playing a key role in the formation of giant planets. 
However some evidences suggest that planetary material might 
have fallen onto the planet-hosting stars (Israelian et al. \cite{israelian01}; 
Murray \& Chaboyer \cite{murray02}) and might be able 
to produce the metallicity excess observed. In this case we might conclude that the 
issue of planetary migration is commonly to be engulfed by the star.
Which of these two scenarios is correct? In other words: is the metallicity
excess observed in planet-hosting stars of primordial origin, or is it due to 
stellar pollution effects?

Asteroseismology consists in measuring properties of 
p-mode oscillations and provides a unique tool for ``drilling'' stellar 
interiors and for testing the internal structure and
chemical composition of bright solar-type stars (e.g. Bouchy \& Carrier 
\cite{bouchycarrier03}; Bedding \& Kjeldsen \cite{bedding03}). 
It also appears as the only means to determine 
the metallicity gradient across the stellar interior in order to solve the question 
of the origin of metallicity excess in planet-hosting stars (Vauclair \cite{vauclair04}; 
Bazot \& Vauclair \cite{bazot04}).  
 
We noted earlier (Bouchy et al., \cite{bouchy04}; Bouchy, \cite{bouchy05}) that 
with a radial velocity precision bellow 1 {\ms}, the HARPS spectrograph (Pepe et al. 
\cite{pepe02}, Mayor et al. \cite{mayor03}) mounted at the 3.6-m telescope offered 
promising perspectives for such asteroseismologic studies.   
We present here results of our asteroseismologic campaign obtained with this instrument 
on HD160691, one of the most metal-rich stars with planetary 
companions. Comparison with stellar structure models and interpretation are presented in 
a companion paper (Bazot et al. \cite{bazot05}).

\section{Observations and data reduction}

HD160691 ($\mu$ Ara, HR6585) is a G3IV-V star with a V magnitude of 5.1 and 
a metallicity [Fe/H] of +0.32. A giant planet (1.7 $M_{\rm Jup}$) in a long period 
(637 days) orbit was detected by Butler et al. (\cite{butler01}). The orbital solution 
was updated by Jones et al. (\cite{jones02}) and Gozdziewski et al. (\cite{gozdziewski03}), 
who found a long term drift possibly 
due to the presence of a second body ($>$ 1.5 $M_{\rm Jup}$, $\sim$1500 days) in the system.   
Bright enough for HARPS, this target is furthermore well sited for an asteroseismology 
campaign (observable more than 10 hours per night). HD160691 was observed over 8 nights 
in June 2004. We took sequences 
of 100-s exposures with a dead time of 31-s in-between. In total 2104 spectra were 
collected with typical night-averaged signal-to-noise ratio (S/N) per pixel in the 
range 95-193 at 550 nm.     

\begin{figure}
\resizebox{8.5cm}{!}{\includegraphics{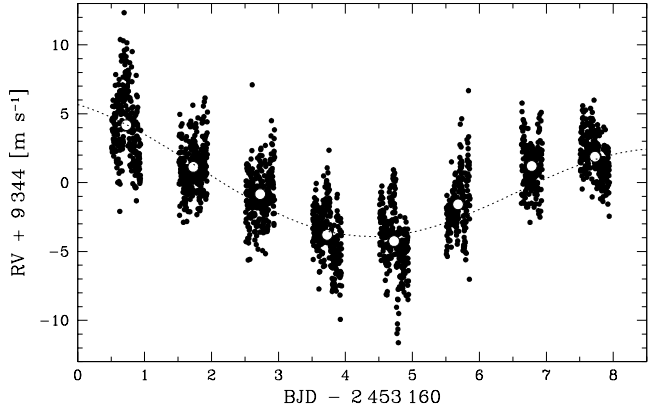}}
\caption{Radial velocity measurements of HD160691. White circles correspond to the night 
average. The dotted curve correspond to the orbital fit of the low-mass-planetary companion 
confirmed thanks to additional measurements by Santos et al. (2004).}
\label{rv}
\end{figure}

\begin{figure}
\resizebox{8.5cm}{!}{\includegraphics{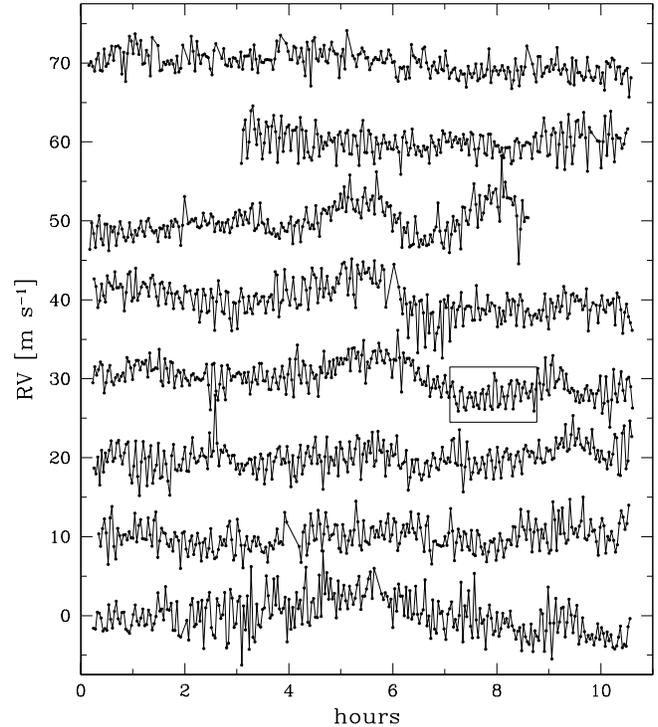}}
\caption{Radial velocity time series of individual nights displayed 
from bottom to top in a relative scale. The box on the fourth night 
is zoomed in Fig.~\ref{rvzoom}.}
\label{rvnights}
\end{figure}

\begin{figure}
\resizebox{8.5cm}{!}{\includegraphics{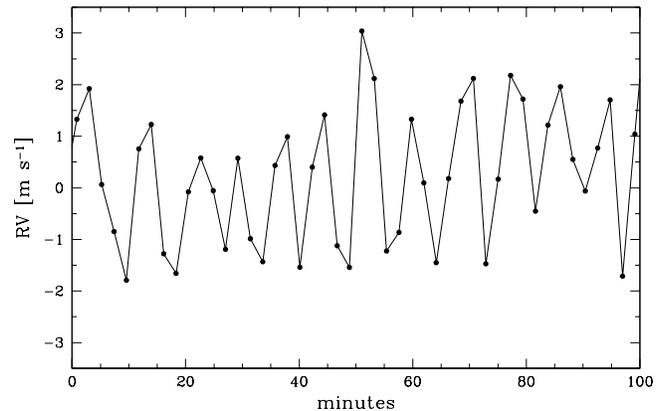}}
\caption{Zoom of radial velocity measurements showing in the time series the presence 
of p-modes with periods around 8 minutes. The semi-amplitude of about 2 {\ms} does not 
represent the individual amplitude of p-modes but comes from the interference of several 
modes.}
\label{rvzoom}
\end{figure}

The spectra obtained were extracted on-line and in real-time using 
the HARPS pipeline. Wavelength calibration was performed with ThAr spectra. 
The radial velocities were obtained by weighted cross-correlation with a numerical 
mask constructed from the Sun spectrum atlas. We also computed off-line the stellar radial 
velocities using the optimum weight procedure as described by Bouchy et al. (\cite{bouchy01}) 
but found no significant gain. 
The journal of the observations is given in Table 1 and the radial velocities 
are presented in Figures~\ref{rv} and \ref{rvnights}. The dispersion of each 
individual nights, in the range 1.5-2.5 {\ms}, is strongly dominated by the acoustic 
modes with period around 8 minutes as shown in Fig.~\ref{rvzoom}.  

The simultaneous ThAr spectrum was used in order to compute the instrumental drift 
using the optimum weight procedure. The spectrograph 
drift was systematically lower than 0.75 {\ms} 
during a night and lower than 2 {\ms} during the whole campaign, illustrating the extreme 
stability of HARPS. Considering this point, we observed without simultaneous Thorium 
during the fourth night in order to construct as more properly as possible 
(without any risk of ThAr contamination) a high S/N combined spectrum for 
spectroscopic analysis of HD160691. 
 
\begin{table}
\caption{Journal of radial velocity measurements. The minimum and maximum S/N of spectra 
obtained during the night is given. The dispersion $\sigma_{RMS}$ is strongly dominated 
by the seismic signal and do not reflect the precision of individual point.}
\begin{center}
\begin{tabular}{ccccc} \hline \hline
Date & Nb spectra & Nb hours & S/N & $\sigma_{RMS}$ [\ms] \\ \hline
2004/06/03 & 279 & 10.32 & 105-181 & 2.49 \\
2004/06/04 & 274 & 10.19 & 65-237 & 1.71 \\
2004/06/05 & 285 & 10.34 & 77-239 & 1.79 \\
2004/06/06 & 286 & 10.38 & 132-235 & 2.03 \\
2004/06/07 & 275 & 10.35 & 43-263 & 2.20 \\
2004/06/08 & 229 & 8.43 & 24-233 & 2.04 \\
2004/06/09 & 202 & 7.43 & 57-147 & 1.65 \\
2004/06/10 & 274 & 10.42 & 32-186 & 1.49 \\ \hline
\end{tabular}
\end{center}
\end{table}

\section{A low-mass-planetary companion}
\label{planet}

Our first surprise was to detect in the Doppler signal an unexpected modulation 
with an semi-amplitude of about 4 {\ms} and a period equal or greater than 8 days. 
Fig.~\ref{rv} clearly shows that we would not have been able to detect 
such a small signal if only one measurement was made each night. Indeed, the semi-amplitude 
of 4 {\ms} corresponds typically to the peak-to-peak variation induced by 
the p-modes in the time series. 

In order to confirm the long period modulation, further observations 
were carried on during two GTO runs (program 073.C-0005) in July and August 2004. 
Considering that the longest periods of p-modes in this star are around 11 minutes 
(see Sect.~\ref{acoustic}), we decided to construct each measurement by averaging 
consecutive independent observations over a period of twice the longest acoustic 
period ($\sim$ 22 minutes). These further measurements allowed us to confirm the 
modulation and determine the period to 9.55 days. This Doppler modulation 
was interpreted as caused by a 14 earth mass planetary companion, the third one 
detected in this exoplanetary system and the lightest planetary companion 
ever observed (Santos et al. \cite{santos04b}). 

\begin{figure}
\resizebox{8.5cm}{!}{\includegraphics{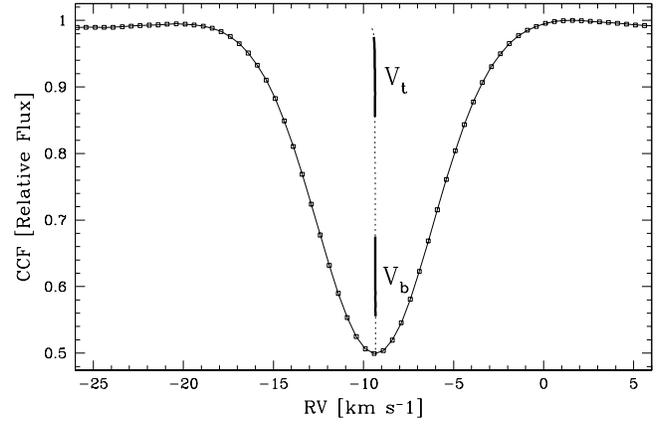}}
\caption{Typical Cross Correlation Function of HD160691 with location of 
the intervals used to compute the bisector velocity span $V_{t}-V_{b}$.}
\label{ccf}
\end{figure}

\begin{figure}
\resizebox{8.5cm}{!}{\includegraphics{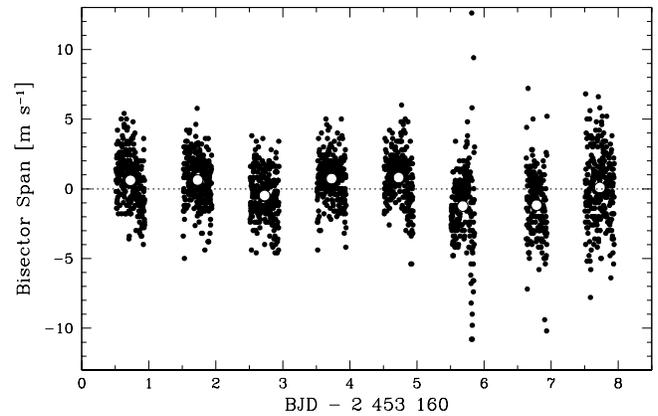}}
\caption{Bisector velocity span of the CCF profile of HD160691. White circles 
correspond to the night average.}
\label{biss}
\end{figure}

In order to verify if this long period modulation was due to a blended binary system 
or a stellar spot, we conducted a bisector analysis of the Cross-Correlation 
Function (CCF) as described in Queloz et al. (\cite{queloz01}) and Santos et al. 
(\cite{santos02}). 
Such an analysis is very efficient to discriminate between radial-velocity 
variations due to changes in the spectral-line profiles from pure Doppler 
variations due to the orbital motion of the star.  
Fig.~\ref{ccf} presents a typical CCF of HD160691. We computed the bisector velocity
of each CCF for 100 different levels (dividing the CCF in 100 slices). We defined the ``bottom'' 
velocity $V_{b}$ and ``top'' velocity $V_{t}$ by averaging the values in the range 
10-35\% and 70-95\% of the CCF depth, respectively. The difference $V_{t}-V_{b}$ 
is equivalent to the bisector velocity span and is used to measure the variations 
of the line profile. The result of our bisector analysis is presented in Fig.~\ref{biss} 
at the same scale as Fig~\ref{rv}. The dispersion of the velocity span on an individual 
night is about 2 {\ms}. No variation appears in phase with the radial-velocity variations 
of period of 9.55 days. 
We checked that no high frequency signals are present in the velocity span 
showing that p-modes induce a pure Doppler variation on the spectral lines. 
The dispersion of the night-averaged-velocity span 
is 0.8 {\ms}. This result indicates that the periodic-radial-velocity change is not 
due to an blended background binary (as HD41004 - see Santos et al. \cite{santos02}) 
nor a dark photospheric stellar spot (as HD166435 - see Queloz et al. \cite{queloz01}).  

\begin{figure}
\resizebox{8.5cm}{!}{\includegraphics{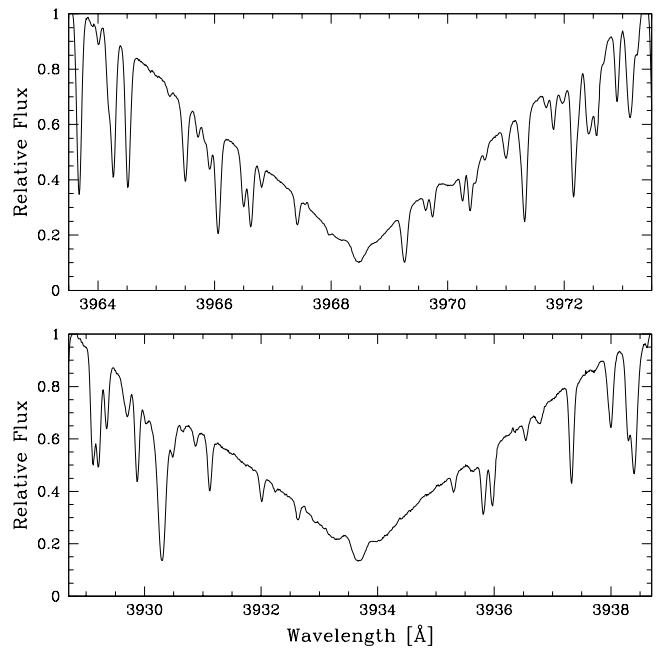}}
\caption{CaII H and K lines for HD160691 showing no chromospheric activity.}
\label{CaII}
\end{figure}

We also checked the magnetic activity of HD160691 by 
computing the chromospheric activity index $S$ based on the relative flux level on 
CaII H and K lines (see Fig.~\ref{CaII}). We measured the flux in two 1 {\AA} pass 
bands centered on the H and K lines normalized by two 20 {\AA} wide sections 
of photospheric flux centered at 3900 and 4000 \AA. The index $S$, computed on 
the 2104 available spectra, has values in the range 0.120-0.125. Once in the Mount 
Wilson scale (Vaughan et al. \cite{vaughan78}), this corresponds to a 
$\log{R'_{HK}}$=-5.034$\pm$0.006 (Noyes et al. \cite{noyes84}). As mentioned 
by Santos et al. (\cite{santos04b}) this value, in close agreement with the one 
obtained by Henry et al. (\cite{henry96}), is typical of an inactive solar type star.

Finally our acoustic spectrum analysis (see Sect.~\ref{acoustic}) allows 
to estimate the rotational period at $\sim$22 days which definitively eliminates 
any stellar activity origin in the 9.55 days signal. 

We noticed that the residuals around the best Keplerian fit 
have a rms of 0.43 {\ms} for the 8 night-averaged measurements made during our 
campaign of asteroseismology (average of more than 200 individual observations) 
and have a rms of 1.3 {\ms} for the other nights (average of 15 observations).
This is due to the presence of low frequency (few hours) modulations in 
the Doppler signal (see Fig.~\ref{rvnights}) with semi-amplitudes of 1-2 {\ms} 
which are not averaged by 15 consecutive independent observations. 
We checked that this noise did not appear in the Thorium signal. 
We also checked that this noise was not introduced by chromatic effect 
on the fiber entrance due to atmospheric-dispersion-correction residuals 
with change of seeing or change of fiber centering. The flux ratio between 
blue and red spectral orders do not present correlation with the low 
frequency modulation seen in the Doppler signal.
The origin of this noise is still unclear, but considering that it does not 
seem to be an instrumental effect and that its amplitude change with time, 
we strongly suspect that it could have a stellar origin (like granulation noise). 

A third exoplanet with low-mass (14 $M_{\oplus}$) and short-period (9.55 days) was not 
completely unexpected and was considered as a possible lucky by-product of this 
asteroseismology campaign. In fact statistics of the exoplanet 
sample show that the probability to find a new exoplanet is higher around 
a planet-hosting star (more than 10\% of known exoplanets are in a multiple system), 
even if the host star is overmetallic. 
We show here that the only way to find such an companion with a semi-amplitude of only 
4.1 {\ms} was to perform such an intensive campaign of radial velocity measurements 
in order to average properly and completely the stellar oscillations. 

\section{Acoustic spectrum analysis}
\label{acoustic}

\begin{figure}
\resizebox{8.5cm}{!}{\includegraphics{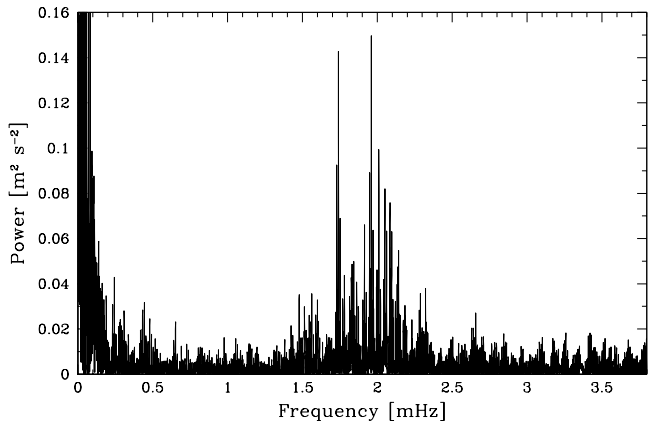}}
\caption{Power spectrum of radial velocity measurements of HD160691.}
\label{tf}
\end{figure}

In order to compute the power spectrum of the velocity time series of Fig.~\ref{rv} 
(corrected from the orbital motion), 
we used the Lomb-Scargle modified algorithm (Lomb \cite{lomb76}, Scargle 
\cite{scargle82}) for unevenly spaced data. The resulting LS periodogram, 
shown in Fig.~\ref{tf}, exhibits a series of peaks between 1.3 and 2.5 mHz modulated 
by a broad envelope, which is the typical signature of solar-like oscillations. 
This signature also appears in the power spectrum of each individual nights. 
The mean white noise level computed in the range 0.7-1.2 mHz is $2.6\times10^{-3}$ 
{\m2s2}. Considering a Gaussian noise, the mean noise level in the amplitude 
spectrum is 4.5 {\cms}. Considering that the time series is based on 2104 measurements, 
the corresponding velocity accuracy thus corresponds 
to 1.17 {\ms}. The uncertainty due to photon noise is estimated to 0.52 {\ms} 
indicating that we are not photon-noise limited. The origin of the external noise 
level of $\sim$1 {\ms} is still unclear. The main instrumental limitation of HARPS 
could comes from the guiding and centering errors on the optical fiber input but tests 
made during commissioning indicated error sources at the level of 0.2 {\ms} (Mayor et al., 
2003). We strongly suspect a stellar origin of this noise (like granulation 
noise).

\begin{figure}
\resizebox{8.5cm}{!}{\includegraphics{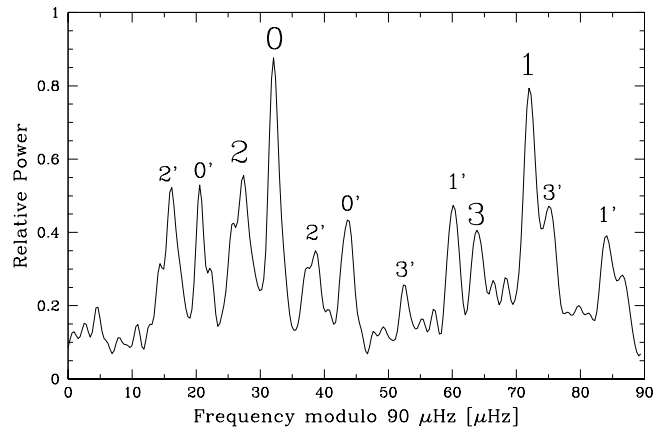}}
\caption{Sum of the echelle diagram showing the modes $l$=0,1,2,3 and their side-lobes.}
\label{sumechel}
\end{figure}

In solar-like stars, p-mode oscillations are expected to produce a
characteristic comb-like structure in the power spectrum with mode
frequencies
$\nu_{n,l,m}$ reasonably well approximated by the simplified asymptotic
relation (Tassoul 1980):
\begin{eqnarray}
\label{eq1}
\nu_{n,l,m} & \approx &
\Delta\nu \cdot (n+\frac{l}{2}+\epsilon) - \delta\nu_{02} \cdot \frac{l(l+1)}{6} - \Omega
\cdot m
\end{eqnarray}

The two quantum numbers $n$ and $l$ correspond to the radial
order and the angular degree of the modes, respectively. For stars 
of which the disk is not resolved, only the lowest-degree modes 
($l\,\leq\,3$) can be detected.
In case of stellar rotation, p-modes need to be
characterized by a third quantum number $m$ called the azimuthal order
($-l\,\leq\,m\,\leq\,l$). 
Quantities $\Delta\nu$, $\delta\nu_{02}$ and $\Omega$ correspond to the large spacing, 
the small spacing and the rotational splitting, respectively.  

\begin{table}
\caption{P-mode frequencies for HD160691 ($\mu$Hz). The frequency resolution 
of the time series is 1.56 $\mu$Hz. Frequencies with label $^a$ correspond 
to peaks with a signal-to-noise ratio in the range 2.5-3 and have to be considered with
caution.}
\begin{center}
\begin{tabular}{lcccc}
\hline\hline
 & $l$ = 0 & $l$ = 1 & $l$ = 2 & $l$ = 3 \\
\hline
n = 13 &  ....  &  ....  & 1381.8$^a$          &  ....  \\
n = 14 &  ....  & 1424.5 &   ....          &  ....  \\ 
n = 15 & 1478.4 & 1513.7 & 1555.1$^a$ / 1557.0$^a$ & 1592.9$^a$ \\
n = 16 & 1562.3 & 1600.2 &   ....          & 1682.5$^a$ \\
n = 17 &  ....  &  ....  & 1731.8 / 1734.9 & 1772.2 \\
n = 18 & 1740.9 & 1780.8 & 1824.2 / 1826.6 & 1862.3 \\
n = 19 & 1831.6 &  ....  & 1914.5          & 1954.7 \\
n = 20 & 1922.0 & 1960.4 & 2005.9 / 2007.8 & 2042.5 \\
n = 21 & 2010.1 & 2051.0 & 2094.9 / 2096.7 & 2132.8 \\
n = 22 & 2101.0 & 2140.8 & 2182.9 / 2184.9 &  ....  \\
n = 23 & 2189.7$^a$ & 2231.0 & 2274.4$^a$ / 2276.0$^a$ & 2315.7$^a$ \\
n = 24 & 2280.3 & 2322.7 &   ....          &  ....  \\
n = 25 &  ....  &  ....  &  ....           &  ....  \\
n = 26 &  ....  & 2504.5$^a$ &   ....          &  ....  \\
\hline
\label{tabfreq}
\end{tabular}\\
\end{center}
\end{table}

\begin{table}
\caption{P-mode amplitudes for HD160691 ({\cms}).}
\begin{center}
\begin{tabular}{lcccc}
\hline\hline
 & $l$ = 0 & $l$ = 1 & $l$ = 2 & $l$ = 3 \\
\hline
n = 13 &  .... &  .... & 11      &  .... \\
n = 14 &  .... & 14    &   ....  &  .... \\ 
n = 15 & 18    & 16    & 10 / 11 & 12    \\
n = 16 & 18    & 17    &   ....  & 12    \\
n = 17 &  .... &  .... & 18 / 15 & 13    \\
n = 18 & 38    & 21    & 16 / 21 & 19    \\
n = 19 & 22    &  .... & 26      & 22    \\
n = 20 & 14    & 39    & 18 / 16 & 19    \\
n = 21 & 32    & 29    & 24 / 25 & 19    \\
n = 22 & 16    & 16    & 15 / 16 &  .... \\
n = 23 & 12    & 15    & 12 / 11 & 11    \\
n = 24 & 16    & 18    &   ....  &  .... \\
n = 25 &  .... &  .... &  ....   &  .... \\
n = 26 &  .... & 12    &   ....  &  .... \\
\hline
\label{tabamp}
\end{tabular}\\
\end{center}
\end{table}

The power spectrum, shown in Fig.~\ref{tf}, presents a clear and unambiguous periodicity 
of 90 $\mu$Hz both on its autocorrelation or in the comb response. 
In order to identify the angular degree $l$ of each mode individually, we 
cut the power spectrum into slices of 90 $\mu$Hz and summed up them. The result, 
which corresponds to the sum of the echelle diagram, is presented on Fig.~\ref{sumechel} 
and allows to identify unambiguously modes $l=0,1,2,3$ and their side-lobes due to the daily 
aliases at $\pm$11.57 $\mu$Hz.  

We selected the strongest peaks with a signal-to-noise ratio greater than 3 in the amplitude 
spectrum and identified them as p-modes with $n$ value based on the asymptotic 
relation (see Eq.~(1)) assuming that the
parameter $\epsilon$ is near the solar value ($\epsilon_{\odot} \sim 1.5$). 
The Frequency and amplitude of these 33 p-modes are listed in Tables~\ref{tabfreq}
and \ref{tabamp}. The amplitude was determined by assuming none of the p-modes 
are resolved and corresponds to the height of the peak in the power spectrum
after quadratic subtraction of the mean noise level. Considering that the frequency 
resolution of our time series is 1.56 $\mu$Hz, we adopted an uncertainty on 
no-resolved oscillation modes of 0.78 $\mu$Hz. Such an uncertainty can be 
considered as conservative in case of infinite time-life modes with high 
signal-to-noise ratio ($\ge$3.5). 
A second inspection of the power spectrum with 
selection of peaks with signal-to-noise ratio in the range 2.5-3 (with amplitude 
in the range 11-13.5 {\cms}) allowed us to propose 10 additional p-modes. 
We are aware that some of these weaker modes could be 
miss-identifications and should be considered with caution and lower confidence. 

\begin{figure}
\resizebox{8.5cm}{!}{\includegraphics{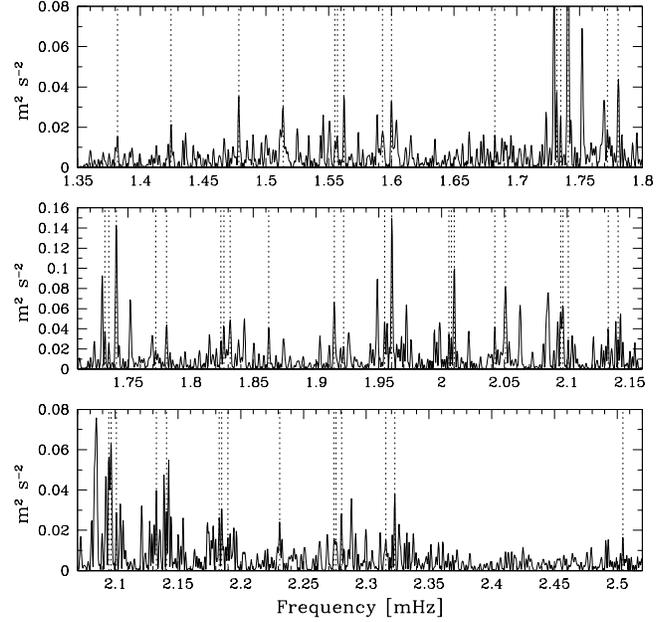}}
\caption{Power spectrum of HD160691 with the identified frequencies in Table~\ref{tabfreq} 
marked by dotted lines.}
\label{tf123}
\end{figure}

\begin{figure}
\resizebox{8.5cm}{!}{\includegraphics{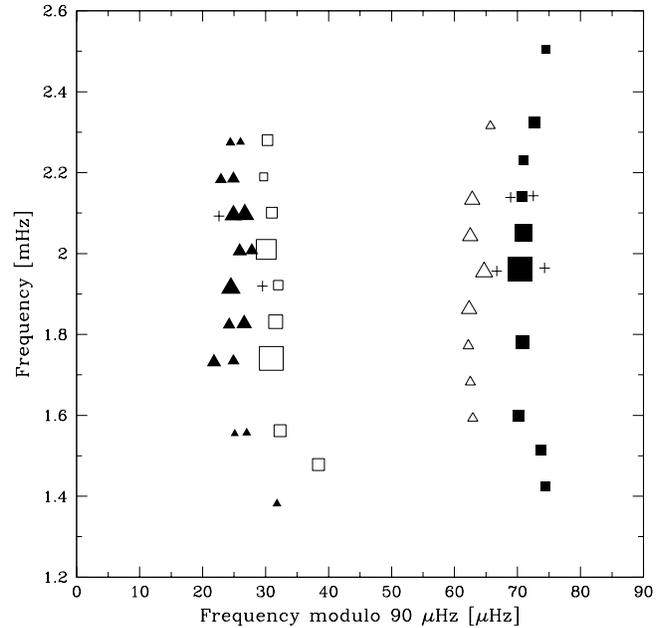}}
\caption{Identified p-modes $l$=0 ($\square$), 
$l$=1 ($\blacksquare$), $l$=2 ($\blacktriangle$), and $l$=3 ($\triangle$) in echelle 
diagram format. The size of the symbols are proportional to the amplitude of the modes. 
Crosses correspond to peaks not identified in the power spectrum.}
\label{echelmode}
\end{figure}

All the identified modes are displayed on the power spectrum in Fig.~\ref{tf123} 
and in the echelle diagram in Fig.~\ref{echelmode}.  
A few peaks with signal-to-noise greater than 3 but not identified as p-modes or 
side-lobes are displayed in Fig.~\ref{echelmode} as crosses. It could 
indicate that some modes are split due to their damping time.  

\begin{figure}
\resizebox{8.5cm}{!}{\includegraphics{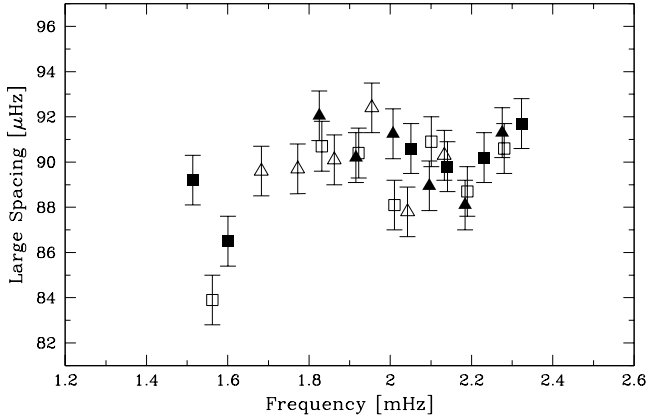}}
\caption{Large spacing $\Delta\nu$ versus frequency for p-modes of degree 
$l$=0 ($\square$), $l$=1 ($\blacksquare$), $l$=2 ($\blacktriangle$) and $l$=3 ($\triangle$).}
\label{largespace}
\end{figure}

\begin{figure}
\resizebox{8.5cm}{!}{\includegraphics{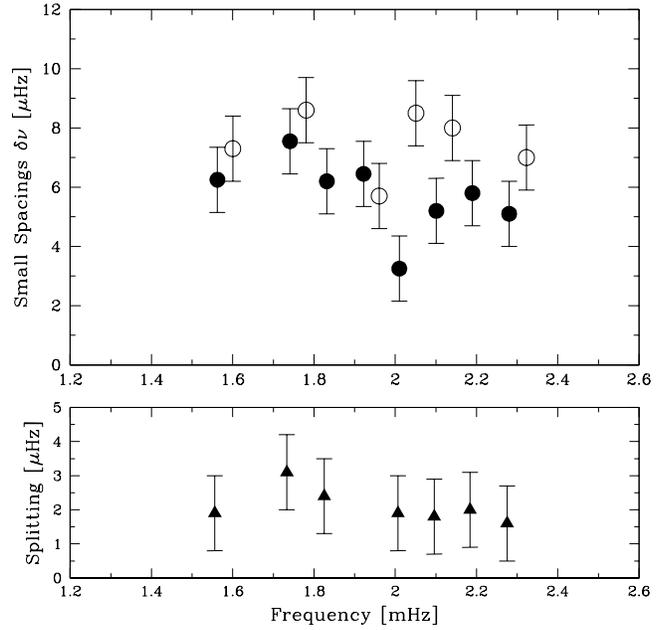}}
\caption{(Top panel) Small spacings $\delta\nu_{0,2}$ ($\bullet$) and $\delta\nu_{1,3}$ 
($\circ$) versus frequencies. (Bottom panel) Splitting of modes $l$=2, $m$=$\pm$2 ($\blacktriangle$).}
\label{smallspace}
\end{figure}

The large spacings $\Delta\nu = \nu_{n+1,l} - \nu_{n,l}$ were computed for 
each pair of consecutive modes with same angular degree. For the $l$=2 modes, 
which are split into 2 components due to the stellar rotation,  
we used the averaged value of the 2 frequencies.  
The result is displayed in Fig.~\ref{largespace}. Neglecting the first three 
points, the residual of the large separations after a third order polynomial fit 
is 1.04 $\mu$Hz RMS. This value indicates that the adopted uncertainty of 
$0.78\,\times\sqrt2\,=\,1.1\,\mu$Hz 
for the large separation is appropriate. 
The small spacings $\delta\nu_{0,2} = \nu_{n+1,0} - \nu_{n,2}$ and  
$\delta\nu_{1,3} = \nu_{n+1,1} - \nu_{n,3}$ are displayed in Fig.~\ref{smallspace}.
The average small spacings $\overline{\delta\nu_{0,2}}$ and  
$\overline{\delta\nu_{1,3}}$ are $5.7\pm0.4$ $\mu$Hz and $7.5\pm0.4$ $\mu$Hz respectively.

The splitting of $l$=2 modes are displayed in Fig.~\ref{smallspace}.
Considering that such a splitting does not appear in modes $l$=1 and the 
fact that the visibilities of modes $m=\pm2$ are higher than modes $m\pm1$, 
we strongly suspect that the observed $l$=2 modes correspond to 
azimuthal order $m=\pm2$. The average splitting of $2.1\pm0.2$ $\mu$Hz leads to 
a rotational splitting $\Omega = 0.53\pm0.05$ $\mu$Hz which corresponds to a 
rotational period of $22\pm2$ days in agreement with the very low magnetic 
activity found in Sect.~\ref{planet}. Modes $l$=3 do not appear to be split. 
It could indicate that we only observe modes (${l=3,m=0}$). An alternative 
is that we observe modes  (${l=3,m=-3}$). In this case modes (${l=3,m=+3}$) 
are expected to be at $\sim$11.1 $\mu$Hz from the $l=1$ modes, hence 
completely merged with the $l$=1 side-lobes.

\section{Conclusions}

Our spectroscopic observations of HD160691 first yield to the discovery of a very 
light planetary companion (14 $M_{\oplus}$) with short period (9.55 days) (Santos et al. 
\cite{santos04b}). Our analysis enforces the validity of this planetary companion 
excluding all sources of ambiguities. 
Our observations of HD160691 also yield a clear detection of p-mode
oscillations. Up to 43 p-modes have been identified in the power spectrum
between 1.3 and 2.5 mHz with an average large spacing of 90 $\mu$Hz, an 
average small spacing of 5.7 $\mu$Hz, a rotational splitting of 0.53 $\mu$Hz and 
an envelope amplitude of about 30 \cms. The identified p-mode frequencies will be 
compared with stellar models in a companion paper (Bazot et al. \cite{bazot05}).

\begin{acknowledgements}
The authors wish to thank R.L. Gilliland for his helpful suggestions.
We are grateful to ESO staff support at the 3.6-m telescope.
S.V. acknowledges a grant from Institut Universitaire de France. 
N.C.S. would like to thank the support from the Swiss National Science Foundation 
and the Portuguese Funda\c{c}\~ao para a Ci\^encia e Tecnologia 
in the form of a scholarship. 
\end{acknowledgements}

\end{document}